# Iron superhydrides FeH$_5$ and FeH$_6$: stability, electronic properties and superconductivity

Alexander G. Kvashnin, [1,2] Ivan A. Kruglov, [2,3] Dmitrii V. Semenok, [1,2] Artem R. Oganov, [1,2,3,4]

[1] Skolkovo Institute of Science and Technology, Skolkovo Innovation Center 143026, 3 Nobel Street, Moscow, Russian Federation
[2] Moscow Institute of Physics and Technology, 141700, 9 Institutsky lane, Dolgoprudny, Russian Federation
[3] Dukhov Research Institute of Automatics (VNIIA), Moscow 127055, Russian Federation
[4] International Center for Materials Discovery, Northwestern Polytechnical University, Xi'an, 710072, China

**Abstract**

Recently a big number of works devoted to search for new hydrides with record high-temperature superconductivity and at the same time the successful synthesis of potential high-$T_C$ superconducting FeH$_5$ was reported. We present a systematic search for stable compounds in the Fe-H system using variable-composition version of the evolutionary algorithm USPEX. All known (FeH, FeH$_3$, FeH$_5$) and several new Fe$_3$H$_5$, Fe$_3$H$_{13}$ and FeH$_6$ iron hydrides were found to be stable, resulting in a very complex phase diagram with rich structural relationships between phases. We calculate electronic properties of two potentially high-$T_C$ FeH$_5$ and FeH$_6$ phases in the pressure range from 150 to 300 GPa. Indeed, hydrogen-rich FeH$_5$ and FeH$_6$ phases were found to be superconducting within Bardeen-Cooper-Schrieffer theory, with $T_C$ values of up to 46 K.

**Introduction**

Theoretical searches for new iron hydrides attracted attention of the scientific community since 1970s when the phase transitions in the Fe-H system were firstly experimentally investigated under pressure. [1,2] There are no stable crystalline iron hydrides at ambient conditions, because hydrogen dissolves in iron in a wide range of pressures and temperatures without the formation of individual phases. [3,4] However such molecules as FeH, FeH$_2$, Fe$_2$H$_4$ and molecular complexes namely FeH$_2$(H$_2$)$_2$, FeH·H$_2$ and FeH$_2$(H$_2$)$_3$ were identified using infrared spectroscopy at low temperatures in inert gas matrix. [5–7] The FeH phase was known as the only one in the Fe-H system for a long time, until the new stable iron hydrides were theoretically predicted in 2012 by Bazhanova et al. [8] using evolutionary algorithm USPEX with fixed-composition searches (restricted to compositions FeH, FeH$_2$, FeH$_3$, FeH$_4$). New FeH$_3$ and FeH$_4$ phases were predicted to be stable at pressure higher than 100 GPa [8] with evidence for likely stability of FeH$_2$ at lower pressures. Recently the structural evolution of FeH$_4$ under pressure was studied as well using fixed-composition search. [9] Several years later in 2014 Pépin et al. [10] have experimentally synthesized the earlier predicted FeH$_3$ at 86 GPa. Besides FeH$_3$ phase another new iron hydride (FeH$_{\sim 2}$) was experimentally found [10], yet precise positions of hydrogen atoms were not established. Recent experimental investigation by Pépin et al. [11] reported about synthesis of FeH$_5$ at pressures above 130 GPa. Its crystal structure was partially determined using available XRD data, but the exact positions of hydrogen atoms were identified using DFT calculations.[3]

Recent exceptional interest was attracted to hydrides due to experimental and theoretical findings of high-$T_C$ superconductivity under high pressures. [12–20] Moreover, recent theoretical investigation of new hydrides in the Ge-H system [21] ($T_C \sim 60$ K), Sn-H [19] ($T_C \sim 100$), MgGeH$_6$ [22] ($T_C \sim 132$ K), H-S [14] ($T_C \sim 200$ K) systems and in Th-H [23] and U-H [20] systems ($T_C \sim 194$ K) at high pressures as well as landmark achievements in experimental synthesis of H$_3$S [17] ($T_C \sim 203$ K), PH$_3$ [24] ($T_C \sim 100$ K), Si$_2$H$_6$ [25] ($T_C \sim 76$ K) inspire exploration of new hydrides. Fresh theoretical work made by Majumdar et al. [26] reported about relatively high possible transition temperature

~ 51 K at 130 GPa in FeH$_5$. All these findings motivate us to study in details the Fe-H system, in particular, stability and superconducting properties of new phases.

Iron is a distinctive element for superconductivity. It was assumed for a long time that magnetism of iron makes superconductivity impossible. However, in 2001 the superconductivity from non-phonon mechanism of interaction was discovered in nonmagnetic ε-Fe with T$_C$ ~ 2.3 K at 15 GPa. [27–29]

Since the unexpected discovery of iron-containing superconducting pnictides in 2008 with potential upper critical magnetic fields of up to 200 T, [30,31] different mechanisms of superconductivity in such materials in relation with magnetism is still under discussion. [32] In this case special interest to FeH$_5$ is caused by the fact that it is the first synthesized superhydride of iron, where superconductivity could come both from electron-phonon coupling and magneto-elastic coupling in the spin-driven scenario [33,34] or/and orbital fluctuation pairing [35,36].

In the last years many developments have been made in order to extend the scope and predictive power of the USPEX code. One of them was the implementation of variable-composition search [37], which allows one to explore the whole compositional space in the studied system in a single calculation. These improvements and previous experimental studies of iron hydrides [10,11] motivated us to perform the evolutionary variable-composition search for new phases in the Fe-H system. The detailed investigation of stability, structural, electronic and superconducting properties of predicted hydrogen-rich phases was carried out, we estimated the possible contribution of electron-phonon interaction to superconductivity of iron hydrides.

## Computational details

We performed variable-composition searches for stable compounds in the Fe-H system at pressure of 0, 50, 100 and 150 GPa using the USPEX [38,39,37] package. The first generation (120 structures) was created using random symmetric generator, while all subsequent generations (100 structures) contained 20% random structures, and 80% created using heredity, softmutation and transmutation operators. Here, evolutionary searches were combined with structure relaxations using density functional theory (DFT) [40,41] within the spin-polarized generalized gradient approximation (Perdew-Burke-Ernzerhof, or PBE functional), [42] and the projector-augmented wave method, [43,44] as implemented in the VASP package. [45–47] Plane wave kinetic energy cutoff was set to 600 eV and the Brillouin zone was sampled by Γ-centered k-points meshes with resolution 2π×0.05 Å$^{-1}$.

By definition, a thermodynamically stable phase has the lowest Gibbs free energy (or, at zero Kelvin, lowest enthalpy) among any phase or phase assemblage of the same composition. Thermodynamic convex hull construction compactly presents information about all possible formation and decomposition reactions; phases that are located on the convex hull are the ones stable at given pressure. Stable structures of elemental Fe and H were taken from USPEX calculations and from Refs. [48–50] and [51], respectively.

Calculations of superconducting T$_C$ were carried out using QUANTUM ESPRESSO package [52]. Phonon frequencies and electron-phonon coupling (EPC) coefficients were computed using density-functional perturbation theory [25], employing plane-wave pseudopotential method and PBE exchange-correlation functional [42]. Convergence tests showed that 70 Ry is a suitable kinetic energy cutoff for the plane wave basis set. We used pseudopotentials with PBE exchange-correlation functional. In our calculations of the electron-phonon coupling (EPC) parameter λ, the first Brillouin zone was sampled using 4×4×1 and 4×4×2 *q*-points mesh, and a denser 16×16×4 and 16×16×8 *k*-points mesh for FeH$_5$ and FeH$_6$, respectively (with Gaussian smearing and σ = 0.05 Ry), which approximates the zero-width limits in the calculation of λ.

Electronic band structures of FeH$_5$ and FeH$_6$ were calculated using both VASP and QE, and demonstrated good consistency. Comparison of the phonon densities of states calculated using finite displacement method (VASP and PHONOPY [54,55]) and density-functional perturbation theory (QE) excellent agreement between these methods.

The superconducting transition temperature $T_C$ was estimated by using two equations: "full" – Allen-Dynes and "short" – modified McMillan equation [56]. The "full" Allen-Dynes equation for calculating $T_C$ has the following form [56]:

$$T_C = \omega_{log} \frac{f_1 f_2}{1.2} \exp\left(\frac{-1.04(1 + \lambda)}{\lambda - \mu^* - 0.62\lambda\mu^*}\right), \quad (1)$$

while the modified McMillan equation has the form as:

$$T_C = \frac{\omega_{log}}{1.2} \exp\left(\frac{-1.04(1 + \lambda)}{\lambda - \mu^* - 0.62\lambda\mu^*}\right). \quad (2)$$

The EPC constant $\lambda$ and logarithmic average frequency $\omega_{log}$ were calculated as:

$$\lambda = \int_{\omega_{min}}^{\omega_{max}} \frac{2 \cdot \alpha^2 F(\omega)}{\omega} d\omega \quad (3)$$

and

$$\omega_{log} = \exp\left(\frac{2}{\lambda} \int_{\omega_{min}}^{\omega_{max}} \frac{d\omega}{\omega} \alpha^2 F(\omega) \ln(\omega)\right), \quad (4)$$

and $\mu^*$ is the Coulomb pseudopotential, for which we used widely accepted lower and upper bound values of 0.10 and 0.15.

Crystal structures of the predicted phases were generated using VESTA software. [57]

## Results and Discussion

In order to predict stable phases in the Fe-H system we performed variable-composition evolutionary searches using the USPEX algorithm [38,39,37] at pressure range from 0 to 150 GPa. Pressure-composition phase was constructed as shown in Fig. 1. There are no stable hydride phases in the pressure range from 0 to 5 GPa, which is in agreement with experimental observations. [3,4] Increase of pressure leads to the formation of $Fm\bar{3}m$-FeH phase which is stable up to 150 GPa. Another experimentally known phase $I4/mmm$-FeH$_2$ is magnetic and stable in a relatively narrow pressure range from 45 to 75 GPa, which is in agreement with literature data. [10] New $Pm\bar{3}m$-Fe$_3$H$_8$ becomes stable from 5 until 75 GPa. It is important that $Pm\bar{3}m$-Fe$_3$H$_8$ is structurally similar to $Pm\bar{3}m$-FeH$_3$ but with one iron and four hydrogen vacancies in the 4×1×1 supercell of FeH$_3$ (see Fig. S6 in Supporting Information). At higher pressures it decomposes to the new $P6_3/mmc$-Fe$_3$H$_5$ and $Pm\bar{3}m$-FeH$_3$. FeH$_3$ is stable from 65 to 150 GPa, which agrees with high-pressure experiments where it was synthesized at 86 GPa. [10] $I4/mmm$-FeH$_5$ phase recently synthesized [11] at pressures above 130 GPa, was found to be thermodynamically stable at pressures from 85 to at least 150 GPa. In the same pressure region the new $I4/mmm$-Fe$_3$H$_{13}$ was theoretically predicted. It important that newly predicted Fe$_3$H$_5$, Fe$_3$H$_8$ and Fe$_3$H$_{13}$ with compositions close to FeH$_2$, FeH$_3$ and FeH$_5$ display rich polysomatism. [58]

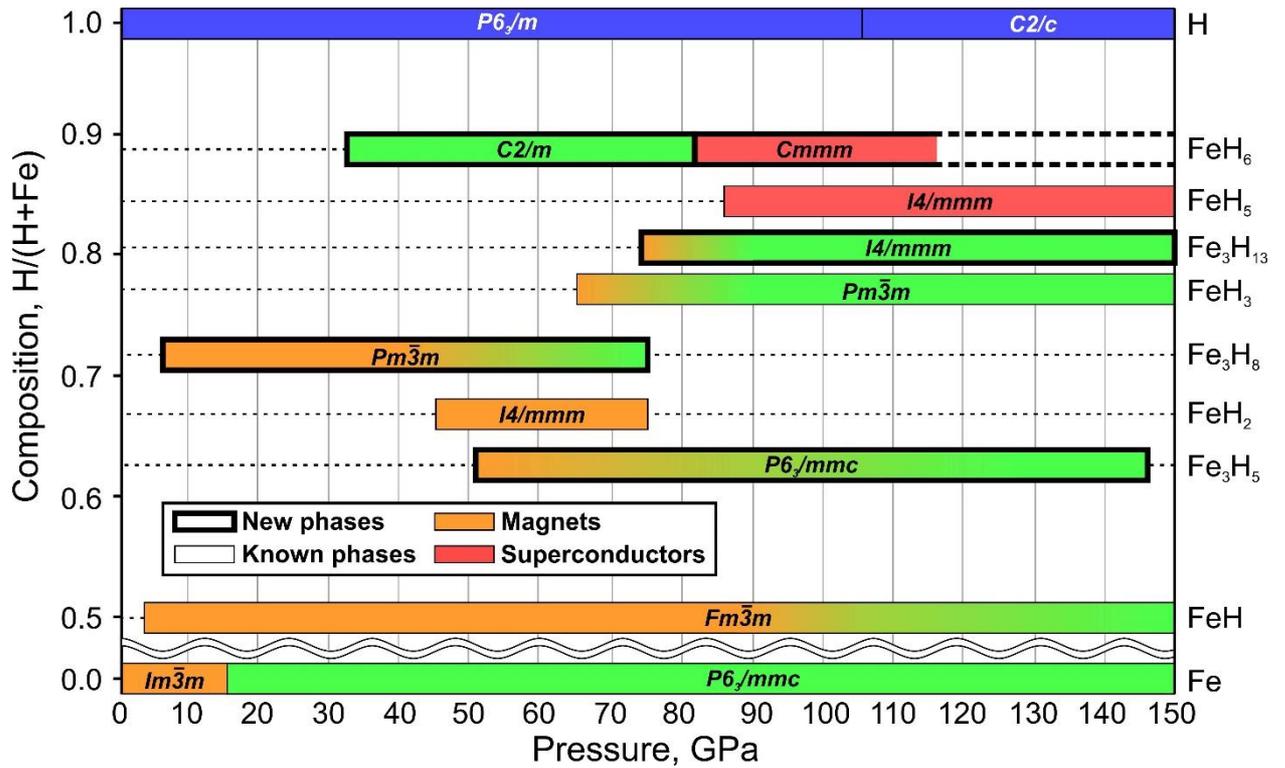

Fig. 1. Pressure-composition phase diagram of the Fe-H system. FeH$_5$ and FeH$_6$ are metallic and superconducting. All shown phases, except hydrogen allotropes are metallic.

We also found new hydrogen-rich $C2/m$-FeH$_6$ phase at pressure higher than 35 GPa to be thermodynamically stable. At ~ 82 GPa $C2/m$-FeH$_6$ transforms to $Cmmm$-FeH$_6$ phase (see Fig. 1), which remains thermodynamically stable at pressures up to 115 GPa (see Supporting Information Fig. S1). At higher pressures (up to 150 GPa) $Cmmm$-FeH$_6$ is only 1.5 meV/atom above decomposition line (see Fig. 2a). Magnetic bcc phase of iron transforms to nonmagnetic hcp ε-Fe at 15 GPa (Fig. 1) in agreement with Refs. 27–29. Most of iron hydrides (FeH, Fe$_3$H$_5$, Fe$_3$H$_8$, FeH$_3$ and Fe$_3$H$_{13}$) are magnetic and become nonmagnetic at high pressures (~ 100 GPa).

Further detailed investigation is devoted to stability, electronic and superconducting properties of Fe-H phases at 150 GPa.

We built the convex hull for the Fe-H system at 150 GPa (see Fig. 2a), which shows that there are 5 stable Fe-H phases namely $Fm\bar{3}m$-FeH, $Pmmm$-Fe$_3$H$_5$, $Pm\bar{3}m$-FeH$_3$, $Immm$-Fe$_3$H$_{13}$ and $I4/mmm$-FeH$_5$. We found a very large number of phases close to convex hull. In the iron-rich part of convex hull (left part of Fig. 2a) the decomposition line has linear character, which corresponds to the formation of solid solutions of hydrogen in iron in a wide range of hydrogen concentrations.[3,4] Hydrogen-rich region (right part of Fig. 2a) with almost parabolic character of decomposition line and phases at this region display rich polysomatism.[58] We found $I4/mmm$-FeH$_2$ to be metastable at 150 GPa by 10 meV/unit. This phase was found in experiments at pressures from 67 GPa up to 86 GPa, but was indicated as FeH$_{\sim 2}$ [10] due to undetermined stoichiometry. Our phase diagram (Fig. 1) explains why FeH$_2$ was not found at higher pressures.

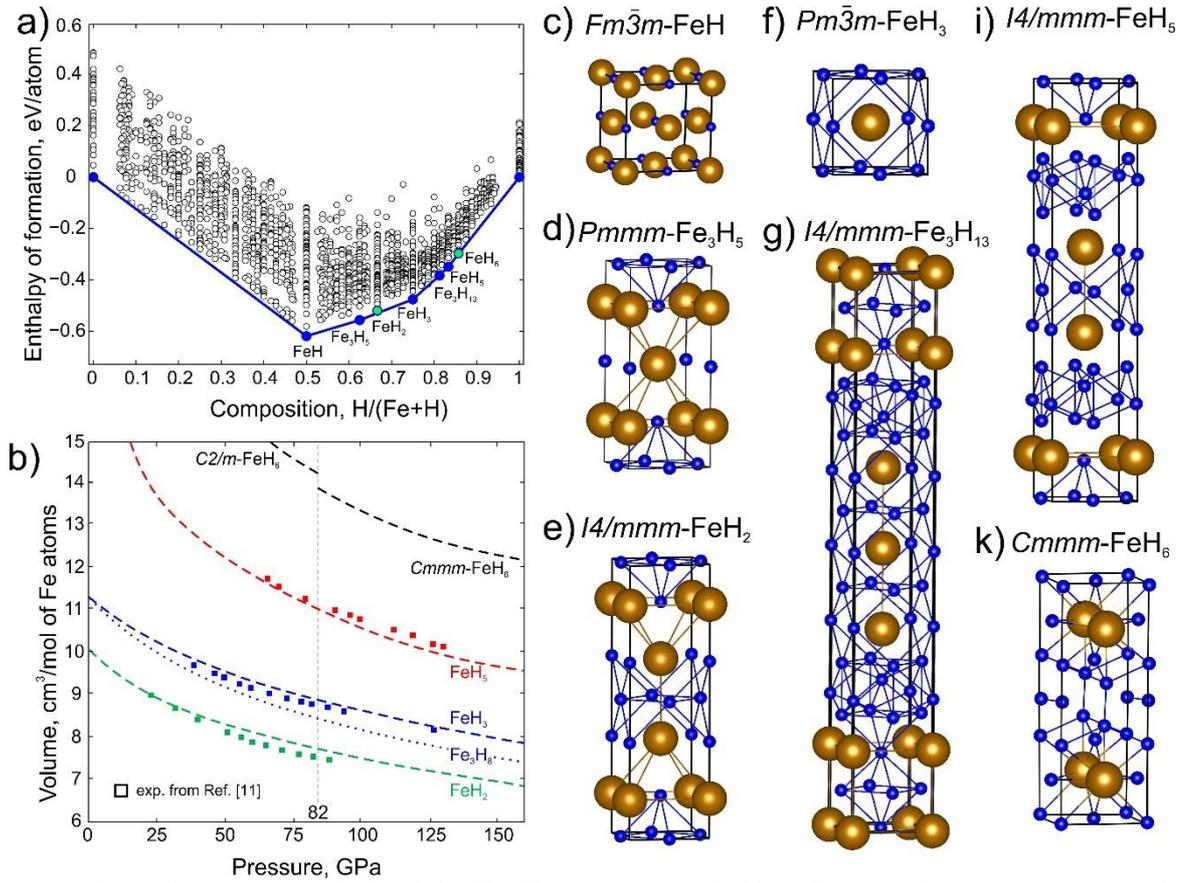

Fig. 2. a) Calculated convex hull of the Fe-H system at 150 GPa. Green points correspond to low-enthalpy metastable FeH$_2$ and FeH$_6$; b) Equations of state of FeH$_2$, Fe$_3$H$_8$, FeH$_3$, FeH$_5$ and FeH$_6$ phases compared with experimental data from Ref. 11; c-g) Crystal structures of predicted stable Fe-H phases. Iron atoms are shown by large brown balls; hydrogen atoms by small blue balls.

Additionally, we calculated equations of state (EOS) of the predicted FeH$_2$, Fe$_3$H$_8$, FeH$_3$, FeH$_5$ and FeH$_6$ phases and compared them with available experimental data from Refs. [10,11] (see Fig. 2b). One can see that theory (dashed lines in Fig. 2b) agrees very well with experimental data (squares in Fig. 2b). EOS of Fe$_3$H$_8$ is close to experimental and calculated data for FeH$_3$ at low pressures (dotted line in Fig. 2b). Comparison of the predicted crystal structures with experiment shows similar positions of iron atoms. Good agreement with all available experimental data lends confidence to our further predictions that the new FeH$_6$ hydride (see black dashed line in Fig. 2b) should be stable at pressures from 50 GPa (in fact, even lower pressures) to 115 GPa (see Supporting Information).

Crystal structures of FeH, FeH$_2$, FeH$_3$, FeH$_5$ and FeH$_6$ are shown in Fig. 2c-g. FeH phase has the well-known rocksalt-type structure (see Fig. 2c). FeH$_2$ has orthorhombic structure alternating FeH$_3$- and FeH-type layers (see Fig. S7 in Supporting Information). FeH$_3$ phase (see Fig. 2e) has the smallest unit cell with only 1 iron atom coordinated by 12 hydrogens; together Fe and H atoms occupy sites on the cubic close packing. Predicted Fe$_3$H$_{13}$ phase has layered structure where each layer is FeH$_3$ phase with the thickness of 4.7 Å (double unit cell) divided by additional hydrogen layer. FeH$_5$ phase is similar to Fe$_3$H$_{13}$ with the only difference in the thickness of FeH$_3$-type layers which equals to 2.36 Å (unit cell of FeH$_3$ phase). Metastable at 150 GPa $Cmmm$-FeH$_6$ phase consists of the FeH$_3$-type layers, similarly to FeH$_5$, but divided by thicker hydrogen layers, than in FeH$_5$ phase (see Fig. 2b) with H$_2$ molecules (H-H distance 0.74 Å) between the layers. Such layered structure with alternation of hydrogen and FeH$_3$-type layers is similar to Na$_3$Cl,[59] which is made of alternating NaCl and Na$_2$ layers. It is currently unknown why such alternating-layer polysomatic compounds become stable under pressure.

Our main interest here is in hydrogen-rich FeH$_5$ and FeH$_6$, and their potential superconductivity in view of recent synthesis and investigations of various superconducting hydrides [12–20,24,25]. It is important to note that both FeH$_5$ and FeH$_6$ remain dynamically stable in the considered pressure region from 150 to 300 GPa, according to phonon calculations (see Supporting Information Fig. S4 and S5).

We have calculated band structures and electronic densities of states of FeH$_5$ and FeH$_6$ phases at 150 GPa (see Fig. 3). The atom-projected band structure shows that contribution to the Fermi level mainly comes from iron atoms (red color), while hydrogen bands lie much deeper in conduction and valence bands (blue color). High peaks of the electronic DOS of FeH$_5$ at -2 eV (see Fig. 3a) can be explained by the presence of flat bands in the direction perpendicular to *c*-axis. Such behavior of the bands indicates the layered structure and weak interaction between layers in FeH$_5$. All other Fe-H phases are metallic with very low DOS at Fermi level (< 0.15 states/eV/unit).

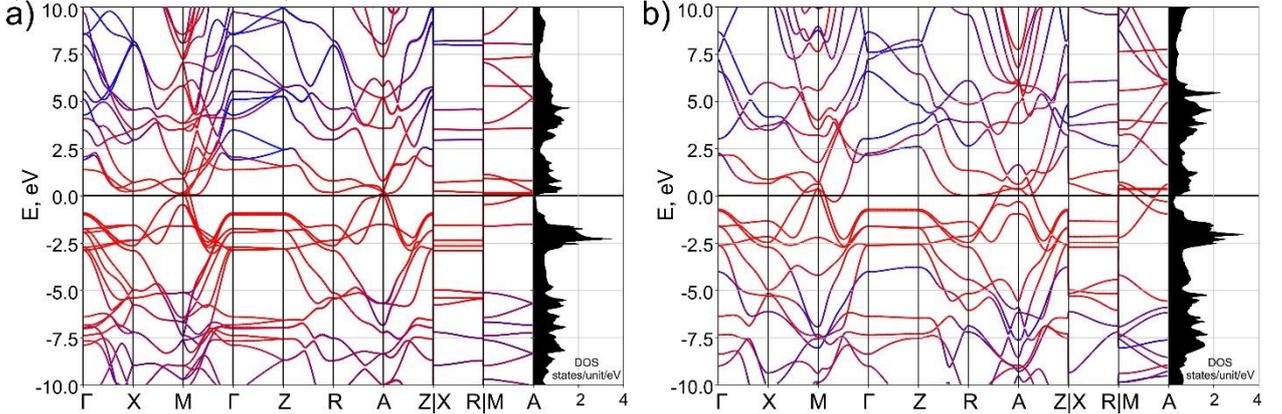

Fig. 3. Electronic band structures and densities of states of a) FeH$_5$ and b) FeH$_6$ at 150 GPa. Red color corresponds to the contribution from Fe, while blue color is for hydrogen atoms.

Similarly, the band structure of FeH$_6$ shows presence of flat bands along the high-symmetry $c^*$ direction of the reciprocal space (see Fig. 3b). We calculated the electronic DOS for both phases at pressures of 150, 200 and 300 GPa. The values of the densities of states at the Fermi level are shown in Table 1. One can see that as pressure increases the density of states slightly increases as well. (see Table 1).

Both crystal structure and electronic properties of FeH$_5$ and FeH$_6$ phase display strong resemblance. We have calculated the electron-phonon coupling (EPC) coefficient $\lambda$, logarithmic frequency $\omega_{\log}$ and T$_C$ using both Allen-Dynes and McMillan formulas [56] as a function of pressure (Table 1). One can see that EPC coefficient for FeH$_5$ phase increases with increasing of pressure due to increasing density of states $N_f$ ($\lambda = N_f \times V_{Coulomb}$). At the same time T$_C$ decreases with pressure, due to decreasing $\omega_{\log}$. Values of T$_C$ for both phases are reasonably high, ~ 43-45 K at pressures 150 GPa, which is in a good agreement with data from Ref. [26] Below 150 GPa FeH$_6$ consistently loses its superconducting properties due to changes in the interlayer interactions. Below 100 GPa we have not found superconductivity in FeH$_6$.

**Table 1.** Predicted superconducting properties of FeH$_5$ and FeH$_6$ phases. T$_C$ values are given for $\mu^* = 0.1$ (0.15).

| Phase | $P$, GPa | $\lambda$ | $N_f$, states/unit/eV | $\omega_{\log}$, K | T$_C$ (Allen-Dynes), K | T$_C$ (McMillan), K |
|---|---|---|---|---|---|---|
| FeH$_5$ | 150 | 0.97 | 0.145 | 642.3 | 45.8 (33.6) | 42.6 (32.3) |
| | 200 | 1.05 | 0.257 | 492.7 | 39.7 (30.6) | 36.6 (28.8) |
| | 300 | 1.26 | 0.318 | 339.5 | 35.7 (28.7) | 32.2 (26.5) |
| FeH$_6$ | 100 | 0.37 | 0.731 | 973.1 | 3.9 (1.2) | 4.0 (1.2) |
| | 150 | 0.92 | 0.436 | 665.6 | 42.9 (31.3) | 40.2 (29.7) |
| | 300 | 0.94 | 0.391 | 549.6 | 37.3 (27.6) | 34.9 (26.1) |

## Conclusions

Using evolutionary crystal structure prediction algorithm USPEX we have uncovered unexpectedly complex chemistry of the Fe-H system at pressure range from 0 to 150 GPa. We confirmed the atomic structure of the experimentally synthesized $FeH_2$, $FeH_3$, $FeH_5$ and predicted new *Pmmm*-$Fe_3H_5$, *Immm*-$Fe_3H_{13}$, *I4/mmm*-$FeH_5$ and *Cmmm*-$FeH_6$ phases to be stable at 150 GPa. All predicted new phases belong to the same polysomatic series formed by the FeH- and $FeH_3$-type blocks. More detailed calculation of *P-V* equation of states allows us to determine the crystal structure of experimentally found $FeH_{\sim2}$ phase as *I4/mmm*-$FeH_2$. Determined crystal structure of hydrogen-rich phases allowed us to perform theoretical calculations of electron DOS and superconducting properties within BCS theory. We showed that both $FeH_5$ and $FeH_6$ demonstrate electronic behavior corresponding to two-dimensional metals. EPC coefficient and DOS at Fermi level increase with decreasing of $\omega_{log}$ in the pressure range from 150 to 300 GPa, which leads to decrease of $T_C$. Thus, $T_C$ values for both *I4/mmm*-$FeH_5$ and *Cmmm*-$FeH_6$ do not exceed 46 K at 150 GPa.

## Acknowledgements

The work was supported by Russian Science Foundation (№ 16-13-10459). Calculations were performed on the Rurik supercomputer at MIPT.

## Reeferences

(1) Stevenson, D. J. Hydrogen in the Earth's Core. *Nature* **1977**, *268* (5616), 130–131.
(2) Wohlfarth, E. P. The Possibility That ϵ-Fe Is a Low Temperature Superconductor. *Phys. Lett. A* **1979**, *75* (1), 141–143.
(3) Armbruster, M. H. The Solubility of Hydrogen at Low Pressure in Iron, Nickel and Certain Steels at 400 to 600°. *J. Am. Chem. Soc.* **1943**, *65* (6), 1043–1054.
(4) Da Silva, J. R. G.; Mclellan, R. B. The Solubility of Hydrogen in Super-Pure-Iron Single Crystals. *J. Common Met.* **1976**, *50* (1), 1–5.
(5) Chertihin, G. V.; Andrews, L. Infrared Spectra of FeH, $FeH_2$, and $FeH_3$ in Solid Argon. *J. Phys. Chem.* **1995**, *99* (32), 12131–12134.
(6) Körsgen, H.; Mürtz, P.; Lipus, K.; Urban, W.; Towle, J. P.; Brown, J. M. The Identification of the $FeH_2$ Radical in the Gas Phase by Infrared Spectroscopy. *J. Chem. Phys.* **1996**, *104* (12), 4859–4861.
(7) Wang, X.; Andrews, L. Infrared Spectra and Theoretical Calculations for Fe, Ru, and Os Metal Hydrides and Dihydrogen Complexes. *J. Phys. Chem. A* **2009**, *113* (3), 551–563.
(8) Bazhanova, Z. G.; Oganov, A. R.; Gianola, O. Fe–C and Fe–H Systems at Pressures of the Earth's Inner Core. *Phys.-Uspekhi* **2012**, *55* (5), 489.
(9) Li, F.; Wang, D.; Du, H.; Zhou, D.; Ma, Y.; Liu, Y. Structural Evolution of FeH 4 under High Pressure. *RSC Adv.* **2017**, *7* (21), 12570–12575.
(10) Pépin, C. M.; Dewaele, A.; Geneste, G.; Loubeyre, P.; Mezouar, M. New Iron Hydrides under High Pressure. *Phys. Rev. Lett.* **2014**, *113* (26), 265504.
(11) Pépin, C. M.; Geneste, G.; Dewaele, A.; Mezouar, M.; Loubeyre, P. Synthesis of $FeH_5$: A Layered Structure with Atomic Hydrogen Slabs. *Science* **2017**, *357* (6349), 382–385.
(12) Gao, G.; Oganov, A. R.; Bergara, A.; Martinez-Canales, M.; Cui, T.; Iitaka, T.; Ma, Y.; Zou, G. Superconducting High Pressure Phase of Germane. *Phys. Rev. Lett.* **2008**, *101* (10), 107002.
(13) Gao, G.; Oganov, A. R.; Li, P.; Li, Z.; Wang, H.; Cui, T.; Ma, Y.; Bergara, A.; Lyakhov, A. O.; Iitaka, T.; et al. High-Pressure Crystal Structures and Superconductivity of Stannane ($SnH_4$). *Proc. Natl. Acad. Sci.* **2010**, *107* (4), 1317–1320.
(14) Duan, D.; Liu, Y.; Tian, F.; Li, D.; Huang, X.; Zhao, Z.; Yu, H.; Liu, B.; Tian, W.; Cui, T. Pressure-Induced Metallization of Dense $(H_2S)_2H_2$ with High-Tc Superconductivity. *Sci. Rep.* **2014**, *4*, 6968.

(15) Hou, P.; Zhao, X.; Tian, F.; Li, D.; Duan, D.; Zhao, Z.; Chu, B.; Liu, B.; Cui, T. High Pressure Structures and Superconductivity of AlH$_3$(H$_2$) Predicted by First Principles. *RSC Adv.* **2015**, *5* (7), 5096–5101.

(16) Li, Y.; Hao, J.; Liu, H.; Tse, J. S.; Wang, Y.; Ma, Y. Pressure-Stabilized Superconductive Yttrium Hydrides. *Sci. Rep.* **2015**, *5*, 09948.

(17) Drozdov, A. P.; Eremets, M. I.; Troyan, I. A.; Ksenofontov, V.; Shylin, S. I. Conventional Superconductivity at 203 Kelvin at High Pressures in the Sulfur Hydride System. *Nature* **2015**, *525* (7567), 73–76.

(18) Goncharov, A. F.; Lobanov, S. S.; Kruglov, I.; Zhao, X.-M.; Chen, X.-J.; Oganov, A. R.; Konôpková, Z.; Prakapenka, V. B. Hydrogen Sulfide at High Pressure: Change in Stoichiometry. *Phys. Rev. B* **2016**, *93* (17), 174105.

(19) Esfahani, M. M. D.; Wang, Z.; Oganov, A. R.; Dong, H.; Zhu, Q.; Wang, S.; Rakitin, M. S.; Zhou, X.-F. Superconductivity of Novel Tin Hydrides (Sn$_n$H$_m$) under Pressure. *Sci. Rep.* **2016**, *6*, srep22873.

(20) Kruglov, I. A.; Kvashnin, A. G.; Goncharov, A. F.; Oganov, A. R.; Lobanov, S.; Holtgrewe, N.; Yanilkin, A. V. High-Temperature Superconductivity of Uranium Hydrides at near-Ambient Conditions. *arXiv:1708.05251* **2017**, *https://arxiv.org/abs/1708.05251*.

(21) Davari Esfahani, M. M.; Oganov, A. R.; Niu, H.; Zhang, J. Superconductivity and Unexpected Chemistry of Germanium Hydrides under Pressure. *Phys. Rev. B* **2017**, *95* (13), 134506.

(22) Ma, Y.; Duan, D.; Shao, Z.; Yu, H.; Liu, H.; Tian, F.; Huang, X.; Li, D.; Liu, B.; Cui, T. Divergent Synthesis Routes and Superconductivity of Ternary Hydride MgSiH$_6$ at High Pressure. *Phys. Rev. B* **2017**, *96* (14), 144518.

(23) Kvashnin, A. G.; Semenok, D. V.; Kruglov, I. A.; Oganov, A. R. High-Temperature Superconductivity in Th-H System at Pressure Conditions. *arXiv:1711.00278* **2017**, http://arxiv.org/abs/1711.00278.

(24) Drozdov, A. P.; Eremets, M. I.; Troyan, I. A. Superconductivity above 100 K in PH3 at High Pressures. *arXiv:1508.06224* **2015**, http://arxiv.org/abs/1508.06224.

(25) Kong, P. P.; Drozdov, A. P.; Eroke, E.; Eremets, M. I. Pressure-Induced Superconductivity above 79 K in Si$_2$H$_6$. In *Book of abstracts of AIRAPT 26 joint with ACHPR 8 & CHPC 19*; Biijing, China, 2017; p 347.

(26) Majumdar, A.; Tse, J. S.; Wu, M.; Yao, Y. Superconductivity in FeH$_5$. *Phys. Rev. B* **2017**, *96* (20), 201107.

(27) Cort, G.; Taylor, R. D.; Willis, J. O. Search for Magnetism in Hcp E-Fe. *J. Appl. Phys.* **1982**, *53* (3), 2064–2065.

(28) Shimizu, K.; Kimura, T.; Furomoto, S.; Takeda, K.; Kontani, K.; Onuki, Y.; Amaya, K. Superconductivity in the Non-Magnetic State of Iron under Pressure. *Nature* **2001**, *412* (6844), 35085536.

(29) Yadav, C. S.; Seyfarth, G.; Pedrazzini, P.; Wilhelm, H.; Černý, R.; Jaccard, D. Effect of Pressure Cycling on Iron: Signatures of an Electronic Instability and Unconventional Superconductivity. *Phys. Rev. B* **2013**, *88* (5), 054110.

(30) Kamihara, Y.; Watanabe, T.; Hirano, M.; Hosono, H. Iron-Based Layered Superconductor La[O$_{1-x}$F$_x$]FeAs (x = 0.05−0.12) with Tc = 26 K. *J. Am. Chem. Soc.* **2008**, *130* (11), 3296–3297.

(31) Zhang, J.; Jiao, L.; Chen, Y.; Yuan, H. Universal Behavior of the Upper Critical Field in Iron-Based Superconductors. *Front. Phys.* **2011**, *6* (4), 463–473.

(32) Hosono, H.; Kuroki, K. Iron-Based Superconductors: Current Status of Materials and Pairing Mechanism. *Phys. C Supercond. Its Appl.* **2015**, *514* (Supplement C), 399–422.

(33) Fernandes, R. M.; Chubukov, A. V.; Schmalian, J. What Drives Nematic Order in Iron-Based Superconductors? *Nat. Phys.* **2014**, *10* (2), 2877.

(34) Eremin, I.; Knolle, J.; Fernandes, R. M.; Schmalian, J.; Chubukov, A. V. Antiferromagnetism in Iron-Based Superconductors: Selection of Magnetic Order and Quasiparticle Interference. *J. Phys. Soc. Jpn.* **2014**, *83* (6), 061015.

(35) Onari, S.; Kontani, H. Self-Consistent Vertex Correction Analysis for Iron-Based Superconductors: Mechanism of Coulomb Interaction-Driven Orbital Fluctuations. *Phys. Rev. Lett.* **2012**, *109* (13), 137001.

(36) Yamada, T.; Ishizuka, J.; Ōno, Y. A High-Tc Mechanism of Iron Pnictide Superconductivity Due to Cooperation of Ferro-Orbital and Antiferromagnetic Fluctuations. *J. Phys. Soc. Jpn.* **2014**, *83* (4), 043704.

(37) Lyakhov, A. O.; Oganov, A. R.; Stokes, H. T.; Zhu, Q. New Developments in Evolutionary Structure Prediction Algorithm USPEX. *Comput. Phys. Commun.* **2013**, *184*, 1172–1182.

(38) Oganov, A. R.; Glass, C. W. Crystal Structure Prediction Using Ab Initio Evolutionary Techniques: Principles and Applications. *J Chem Phys* **2006**, *124*, 244704.

(39) Oganov, A. R.; Lyakhov, A. O.; Valle, M. How Evolutionary Crystal Structure Prediction Works—and Why. *Acc. Chem. Res.* **2011**, *44*, 227–237.

(40) Hohenberg, P.; Kohn, W. Inhomogeneous Electron Gas. *Phys Rev* **1964**, *136* (3B), B864–B871.

(41) Kohn, W.; Sham, L. J. Self-Consistent Equations Including Exchange and Correlation Effects. *Phys Rev* **1965**, *140* (4), A1133–A1138.

(42) Perdew, J. P.; Burke, K.; Ernzerhof, M. Generalized Gradient Approximation Made Simple. *Phys. Rev. Lett.* **1996**, *77* (18), 3865–3868.

(43) Blöchl, P. E. Projector Augmented-Wave Method. *Phys. Rev. B* **1994**, *50* (24), 17953–17979.

(44) Kresse, G.; Joubert, D. From Ultrasoft Pseudopotentials to the Projector Augmented-Wave Method. *Phys. Rev. B* **1999**, *59* (3), 1758–1775.

(45) Kresse, G.; Furthmüller, J. Efficient Iterative Schemes for Ab Initio Total-Energy Calculations Using a Plane-Wave Basis Set. *Phys. Rev. B* **1996**, *54*, 11169–11186.

(46) Kresse, G.; Hafner, J. Ab Initio Molecular Dynamics for Liquid Metals. *Phys. Rev. B* **1993**, *47*, 558–561.

(47) Kresse, G.; Hafner, J. Ab Initio Molecular-Dynamics Simulation of the Liquid-Metal Amorphous-Semiconductor Transition in Germanium. *Phys. Rev. B* **1994**, *49*, 14251–14269.

(48) Saxena, S. K.; Dubrovinsky, L. S.; Häggkvist, P.; Cerenius, Y.; Shen, G.; Mao, H. K. Synchrotron X-Ray Study of Iron at High Pressure and Temperature. *Science* **1995**, *269* (5231), 1703–1704.

(49) Belonoshko, A. B.; Ahuja, R.; Johansson, B. Stability of the Body-Centred-Cubic Phase of Iron in the Earth's Inner Core. *Nature* **2003**, *424* (6952), 1032–1034.

(50) Tateno, S.; Hirose, K.; Ohishi, Y.; Tatsumi, Y. The Structure of Iron in Earth's Inner Core. *Science* **2010**, *330* (6002), 359–361.

(51) Pickard, C. J.; Needs, R. J. Structure of Phase III of Solid Hydrogen. *Nat. Phys.* **2007**, *3* (7), 473–476.

(52) Giannozzi, P.; Baroni, S.; Bonini, N.; Calandra, M.; Car, R.; Cavazzoni, C.; Ceresoli, D.; Chiarotti, G. L.; Cococcioni, M.; Dabo, I.; et al. QUANTUM ESPRESSO: A Modular and Open-Source Software Project for Quantum Simulations of Materials. *J. Phys. Condens. Matter* **2009**, *21*, 395502.

(53) Baroni, S.; de Gironcoli, S.; Dal Corso, A.; Giannozzi, P. Phonons and Related Crystal Properties from Density-Functional Perturbation Theory. *Rev. Mod. Phys.* **2001**, *73* (2), 515–562.

(54) Togo, A.; Tanaka, I. First Principles Phonon Calculations in Materials Science. *Scr. Mater.* **2015**, *108*, 1–5.

(55) Togo, A.; Oba, F.; Tanaka, I. First-Principles Calculations of the Ferroelastic Transition between Rutile-Type and $CaCl_2$-Type $SiO_2$ at High Pressures. *Phys. Rev. B* **2008**, *78*, 134106.

(56) Allen, P. B.; Dynes, R. C. Transition Temperature of Strong-Coupled Superconductors Reanalyzed. *Phys. Rev. B* **1975**, *12* (3), 905–922.

(57) Momma, K.; Izumi, F. VESTA 3 for Three-Dimensional Visualization of Crystal,

Volumetric and Morphology Data. *J. Appl. Crystallogr.* **2011**, *44*, 1272–1276.
(58) Veblen, D. R. Polysomatism and Polysomatic Series: A Review and Applications. *Am. Minaralogist* **1991**, *76*, 801–826.
(59) Zhang, W.; Oganov, A. R.; Goncharov, A. F.; Zhu, Q.; Boulfelfel, S. E.; Lyakhov, A. O.; Somayazulu, M.; Prakapenka, V. B. Unexpected Stable Stoichiometries of Sodium Chlorides. *Science* **2013**, *342* (6165 DOI-10.1126/science.1244989), 1502–1505.

# Supporting information

# Iron superhydrides FeH$_5$ and FeH$_6$: stability, electronic properties and superconductivity

*Alexander G. Kvashnin,* [1,2] *Ivan A. Kruglov,* [2,3] *Dmitrii V. Semenok,* [1,2] *Artem R. Oganov,* [1,2,3,4]

[1] Skolkovo Institute of Science and Technology, Skolkovo Innovation Center 143026, 3 Nobel Street, Moscow, Russian Federation

[2] Moscow Institute of Physics and Technology, 141700, 9 Institutsky lane, Dolgoprudny, Russian Federation

[3] Dukhov Research Institute of Automatics (VNIIA), Moscow 127055, Russian Federation

[4] International Center for Materials Discovery, Northwestern Polytechnical University, Xi'an, 710072, China

## Equations for calculation $T_C$

The "full" Allen-Dynes equation for calculating $T_C$ has the following form [1]:

$$T_C = \omega_{log} \frac{f_1 f_2}{1.2} exp\left(\frac{-1.04(1+\lambda)}{\lambda - \mu^* - 0.62\lambda\mu^*}\right), \quad (S5)$$

while the modified McMillan equation has the form as:

$$T_C = \frac{\omega_{log}}{1.2} exp\left(\frac{-1.04(1+\lambda)}{\lambda - \mu^* - 0.62\lambda\mu^*}\right), \quad (S6)$$

where

$$f_1 = \left(1 + \left(\frac{\lambda}{\Lambda_1}\right)^{\frac{3}{2}}\right)^{\frac{1}{3}}, \quad f_2 = 1 + \frac{\lambda^2}{\lambda^2 + \Lambda_2^2} \cdot \left(\frac{\omega_2}{\omega_{log}} - 1\right) \quad (S7)$$

and

$$\Lambda_1 = 2.46(1 + 3.8\mu^*), \quad \Lambda_2 = 1.82(1 + 6.3\mu^*),$$

$$\omega_2 = \sqrt{\frac{1}{\lambda} \int_{\omega_{min}}^{\omega_{max}} \left[\frac{2 \cdot a^2 F(\omega)}{\omega}\right] \omega^2 d\omega} \quad (S8)$$

$f_1$, $f_2$, $\Lambda_1$, $\Lambda_2$ are functions and correlation factors, used in Allen-Dynes paper for calculating the critical temperature in the case of the inapplicability of the McMillan approximation for large EPC parameters $\lambda > 2$, $\omega_2$ is square root of the mean squared frequency. Minimal ($\omega_{min} > 0$) and maximal ($\omega_{max}$) frequencies were determined as the extreme zero values of the Eliashberg function, where $a^2F(\omega) < 10^{-6}$.

## Stability ranges of new iron polyhydrides

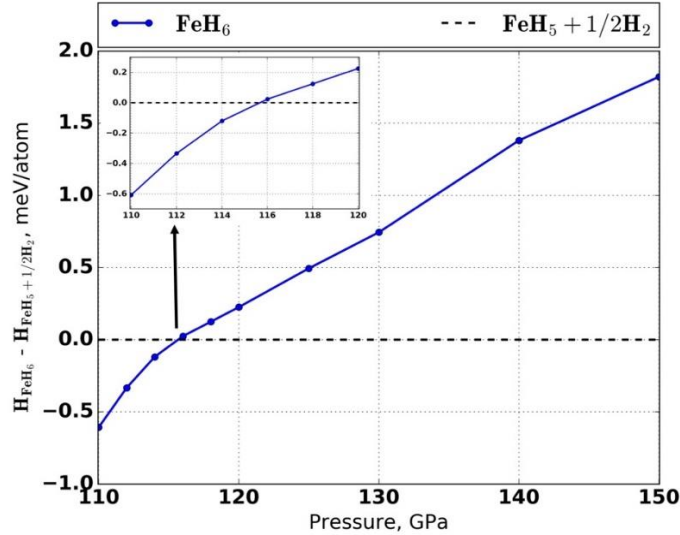

**Fig. S1.** Decomposition enthalpy curve of $FeH_6$ into $FeH_5$ and $H_2$.

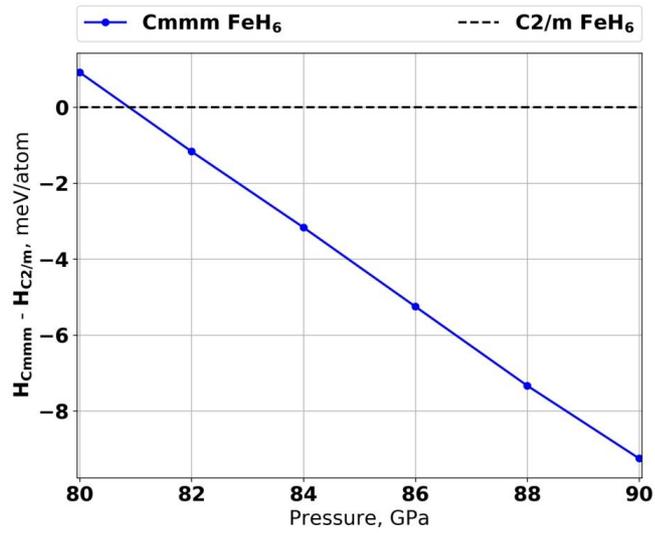

**Fig. S2.** Enthalpy difference between $Cmmm$ and $C2/m$-FeH$_6$ phases.

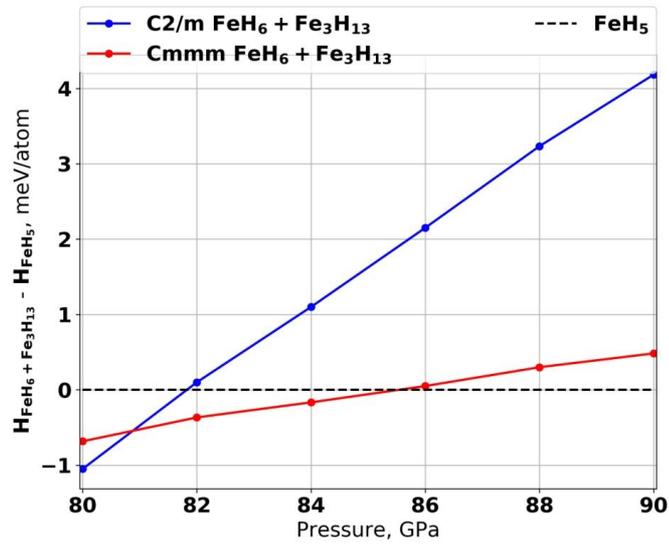

**Fig. S3.** Enthalpy of formation of FeH$_5$ formation from FeH$_6$ and Fe$_3$H$_{13}$

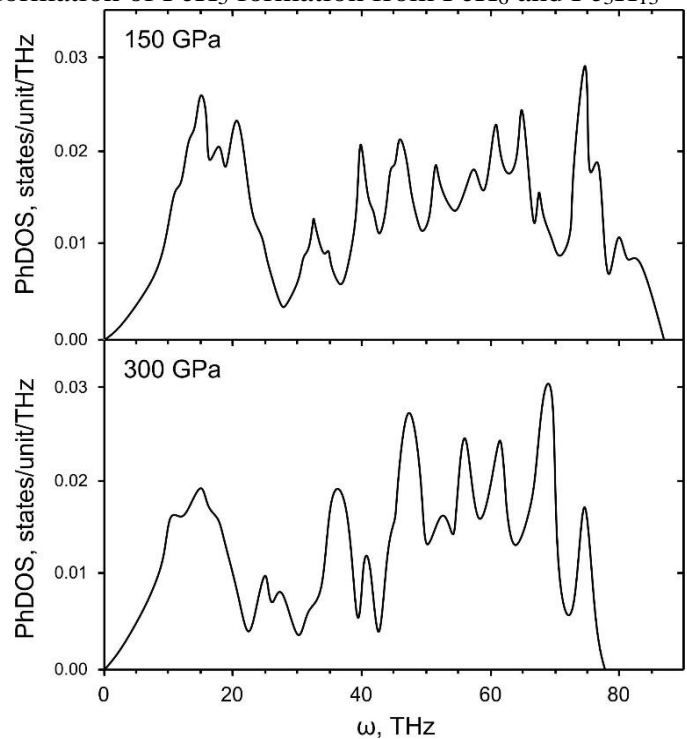

**Fig. S4.** Phonon DOS of FeH$_5$ at 150 and 300 GPa.

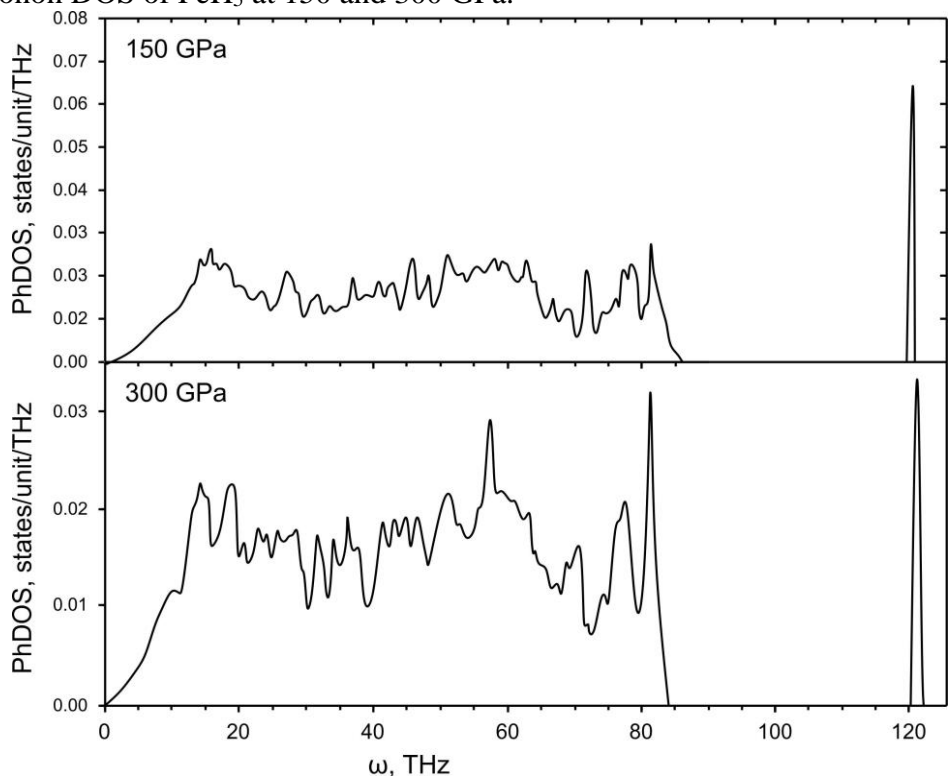

**Fig. S5.** Phonon DOS of FeH$_6$ at 150 and 300 GPa.

## Crystal structures of FeH, FeH$_2$, FeH$_3$ and Fe$_3$H$_8$ phases

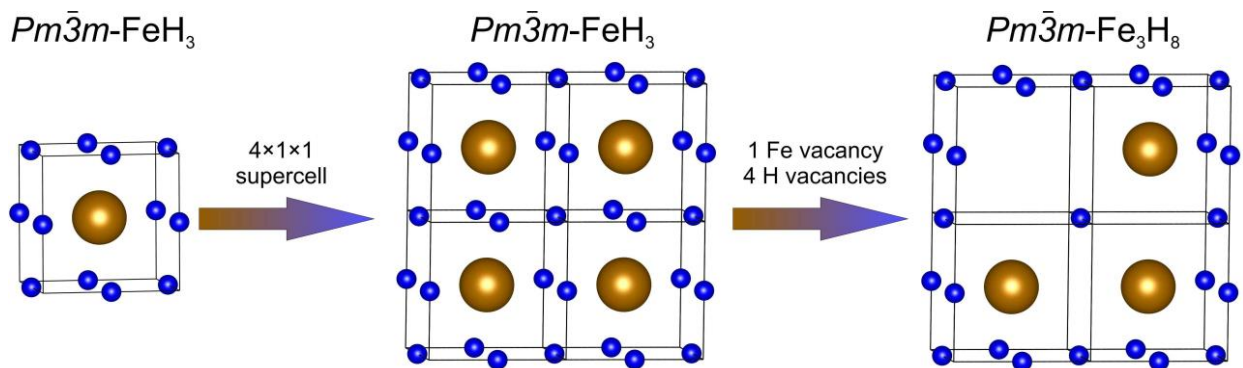

**Fig. S6.** Similarity between crystal structure of FeH$_3$ and Fe$_3$H$_8$. One iron and four hydrogen vacancies transform 4×1×1 supercell of FeH$_3$ to Fe$_3$H$_8$.

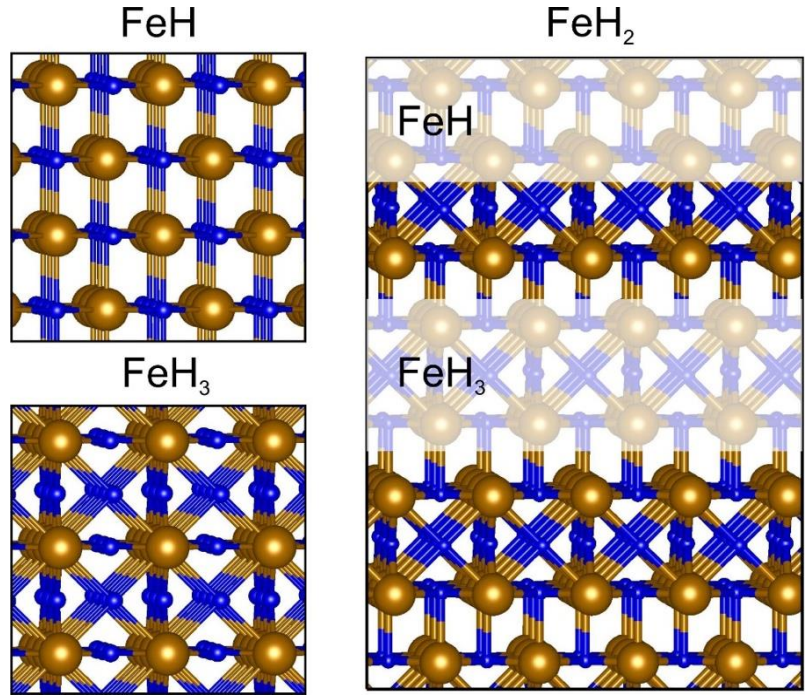

**Fig. S7.** Crystal structures of Fe-H phase showing the alternation of FeH- and FeH$_3$-type layers along *c*-axis in FeH$_2$ phase.

## Electronic properties of Fe-H phases

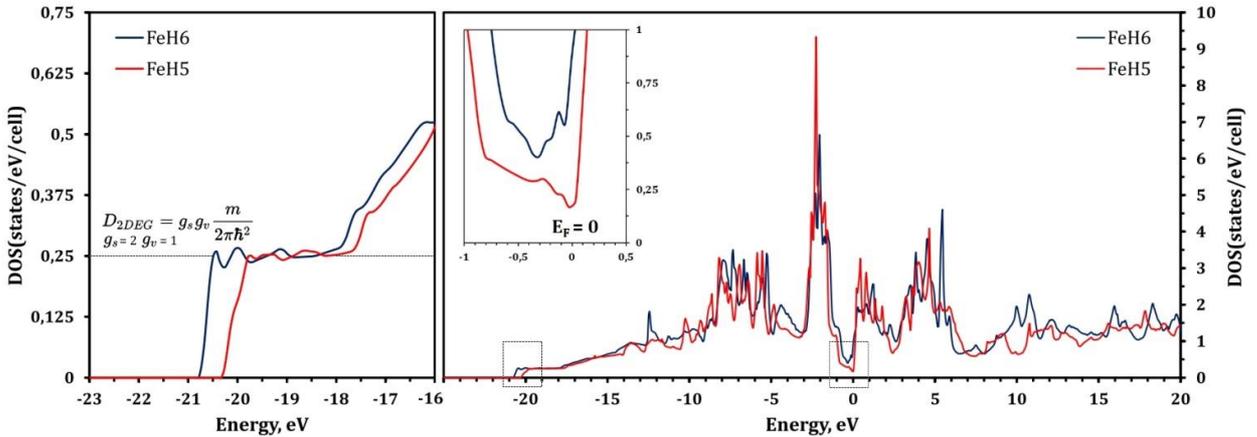

**Fig. S8.** Electronic density of states for $Fm\bar{3}m$-FeH$_5$ and $Cmmm$-FeH$_6$ phases at 150 GPa.

As can be seen from the Fig. S6 the DOS(E) functions can be considered as combinations of 2 contributions: interlayer electronic density of states (3D) and intralayer density of states (2D) localized between the iron atoms in the layers. The latter contribution numerically equals to 0.25 states/eV/cell (the cell parameters of 2D FeH$_5$ sublattice at 150 GPa are a = b = 2.39 Å) and clearly manifests itself near Fermi level (see inset of right panel of Fig. S8) and in the region from -17.5 to -20 eV, where the 3D contribution is almost zero (inset of left panel of Fig. S8). Two-dimensional density of states does not depend on energy and identifying by the characteristic step (see inset of left panel of Fig. S8). Moreover, the intralayer conductivity causes the metallic properties of FeH$_5$, FeH$_6$ at 150 GPa.

## Eliashberg spectral functions for FeH$_5$, FeH$_6$ phases at different pressures

In these all cases the stretch contribution is weak (from 120 to 140 THz) and has not shown on the diagram for clarity

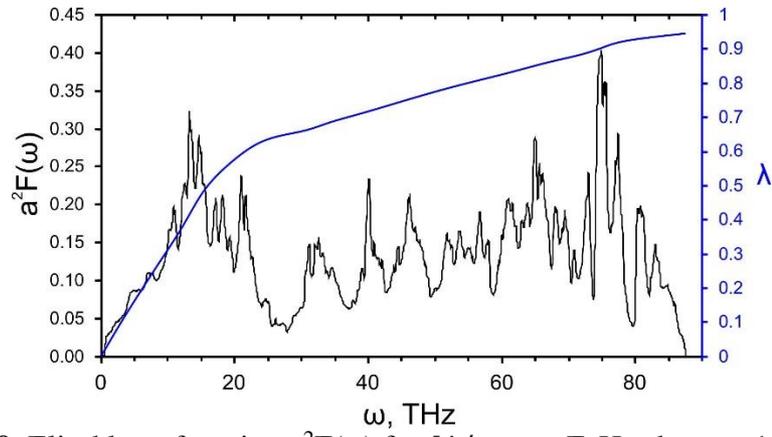

**Fig. S9.** Eliashberg function $a^2F(\omega)$ for $I4/mmm$-FeH$_5$ phase at 150 GPa

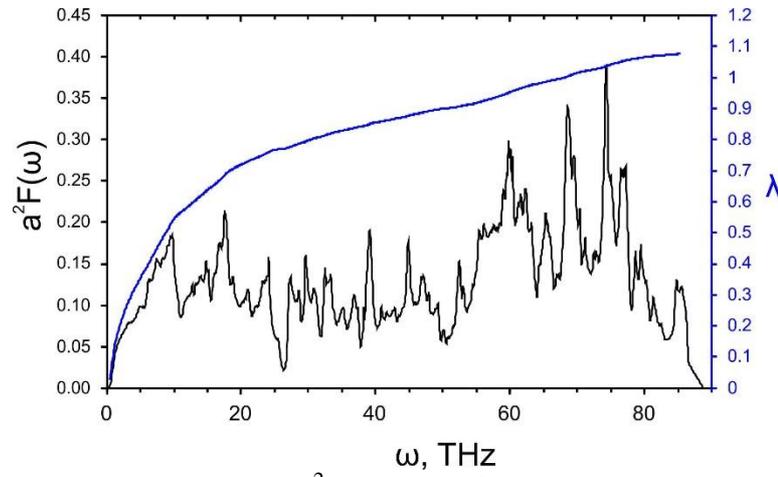

**Fig. S10.** Eliashberg function $a^2F(\omega)$ for $I4/mmm$-FeH$_5$ phase at 200 GPa

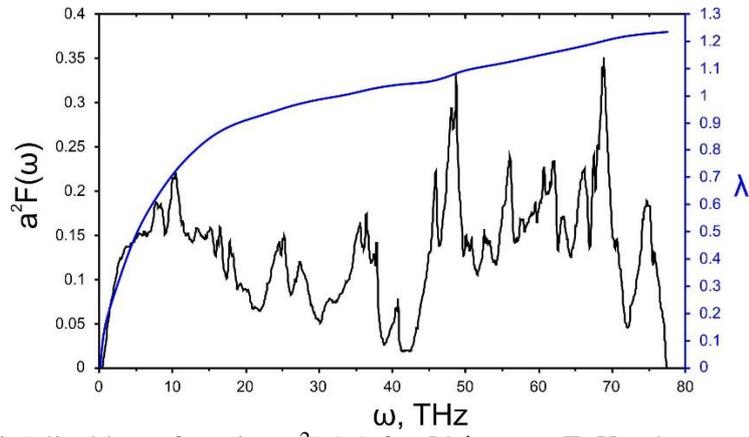

**Fig. S11.** Eliashberg function a$^2$F(ω) for $I4/mmm$-FeH$_5$ phase at 300 GPa

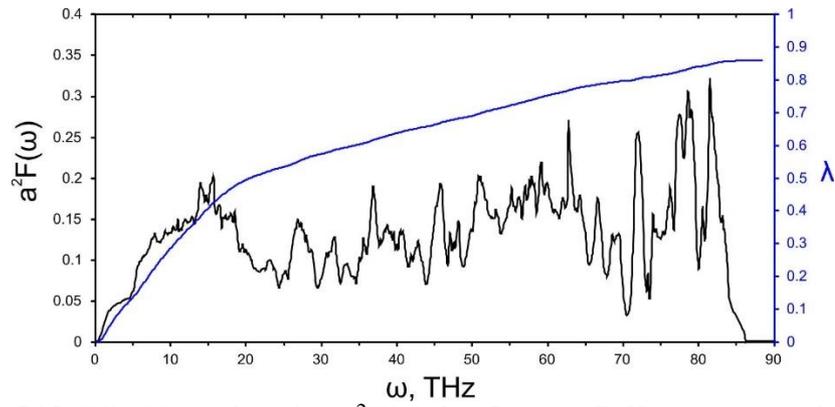

**Fig. S12.** Eliashberg function a$^2$F(ω) for $Cmmm$-FeH$_6$ phase at 150 GPa

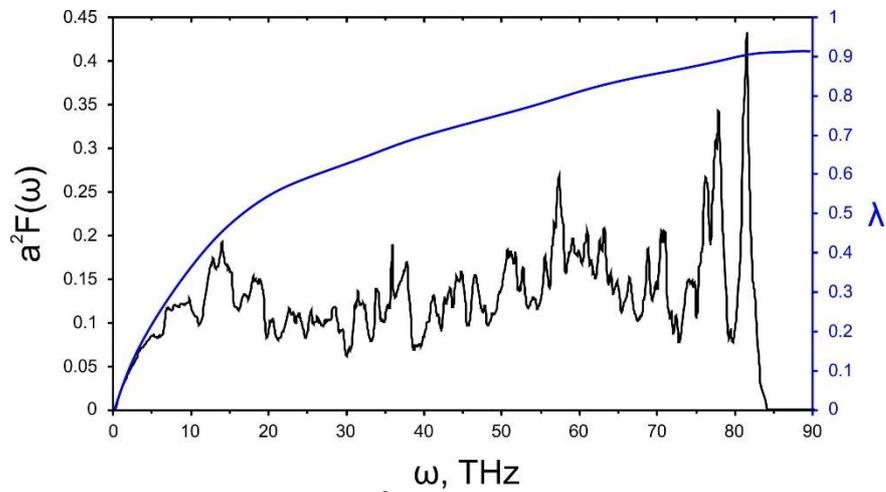

**Fig. S13.** Eliashberg function a$^2$F(ω) for $Cmmm$-FeH$_6$ phase at 300 GPa

## References

(1) Allen, P. B.; Dynes, R. C. Transition Temperature of Strong-Coupled Superconductors Reanalyzed. *Phys. Rev. B* **1975**, *12*, 905–922.